\documentclass[10pt,journal]{IEEEtran}
\IEEEoverridecommandlockouts
\usepackage{scrextend}
\changefontsizes[11.1pt]{9.5pt}

\usepackage{booktabs} 
\usepackage[dvipsnames]{xcolor}
\usepackage{amsmath}

\usepackage[linesnumbered,ruled,vlined]{algorithm2e}
\usepackage{adjustbox}
\usepackage{graphicx}
\usepackage{soul,xcolor}
\usepackage{graphicx}
\usepackage{multirow}
\usepackage{epsfig}
\usepackage{rotating}
\usepackage{array}
\usepackage{enumitem}
\usepackage[normalem]{ulem}
\usepackage{dblfloatfix}
\usepackage[font=small,labelfont=bf,skip=3pt]{caption}
\usepackage{graphics}
\usepackage{ragged2e}
\usepackage[utf8]{inputenc}
\usepackage[T1]{fontenc}
\usepackage{gensymb}
\usepackage{amssymb}
\usepackage{pifont}

\usepackage{titlesec}
\titlespacing\section{0pt}{4pt plus 1pt minus 1pt}{4pt plus 1pt minus 1pt}
\titlespacing\subsection{0pt}{4pt plus 1pt minus 1pt}{4pt plus 1pt minus 1pt}
\titlespacing\subsubsection{0pt}{4pt plus 1pt minus 1pt}{4pt plus 1pt minus 1pt}

\definecolor{OliveGreen}{rgb}{0,0.6,0}
\definecolor{skyblue}{rgb}{0.06, 0.75, 0.99}

\newcommand{\new}[1]{\textcolor{black}{#1}}

\newcommand{\rev}[1]{\textcolor{black}{#1}}

\setstcolor{black}

\def\BibTeX{{\rm B\kern-.05em{\sc i\kern-.025em b}\kern-.08em
    T\kern-.1667em\lower.7ex\hbox{E}\kern-.125emX}}

\begin{document}

\title{DAS: Dynamic Adaptive Scheduling for \\ Energy-Efficient Heterogeneous SoCs 

\author{A.~Alper~Goksoy, Anish~Krishnakumar, Md~Sahil~Hassan, Allen~J.~Farcas, \\ Ali~Akoglu, Radu~Marculescu, and Umit~Y.~Ogras}

\thanks{
\indent This material is based on research sponsored by Air Force Research Laboratory (AFRL) and Defense Advanced Research Projects Agency (DARPA) under agreement number FA8650-18-2-7860. The U.S. Government is authorized to reproduce and distribute reprints for Governmental purposes notwithstanding any copyright notation thereon. The views and conclusion contained herein are those of the authors and should not be interpreted as necessarily representing the official policies or endorsements, either expressed or implied, of Air Force Research Laboratory (AFRL) and Defense Advanced Research Projects Agency (DARPA) or the U.S. Government. \newline
\indent A.A.Goksoy, A.Krishnakumar, and U.Y.Ogras are with the Dept. of Electrical and Computer Engineering, University of Wisconsin-Madison, Madison, WI 53705 USA. (Corresponding author: A. Alper Goksoy) \newline 
E-mail: \{agoksoy, anish.n.krishnakumar, uogras\}@wisc.edu \newline
\indent M.S.Hassan and A.Akoglu are with the Dept. of Electrical and Computer Engineering, University of Arizona, Tucson, AZ 85719 USA. \newline E-mail: \{sahilhassan, akoglu\}@email.arizona.edu \newline
\indent A.J.Farcas and R.Marculescu are with the Dept. of Electrical and Computer Engineering, University of Texas at Austin, Austin, TX 78712 USA. \newline
E-mail: \{allen.farcas, radum\}@utexas.edu \newline

}

}
\IEEEaftertitletext{\vspace{-1\baselineskip}}
\maketitle
\begin{abstract}

Domain-specific systems-on-chip (DSSoCs) aim at bridging the gap between application-specific integrated circuits (ASICs) and general-purpose processors. 
Traditional operating system (OS) schedulers can undermine the potential of DSSoCs since their execution times can be orders of magnitude larger than the execution time of the task itself. 
To address this problem, we propose a dynamic adaptive scheduling (DAS) framework that combines the benefits of a fast (low-overhead) scheduler and a slow (sophisticated, high-performance but high-overhead) scheduler.
Experiments with five real-world streaming applications show that DAS consistently outperforms both the fast and slow schedulers. For 40 different workloads, DAS achieves on average 1.29$\times$ speedup and 45\% lower EDP compared to the sophisticated scheduler at low data rates and 1.28$\times$ speedup and 37\% lower EDP than the fast scheduler when the workload complexity increases.

\end{abstract}


\begin{IEEEkeywords}
Domain-specific SoC (DSSoC), machine learning, scheduling, runtime classification, policy switching.\end{IEEEkeywords} 



\section{Introduction}
\label{sec:introduction}
Heterogeneous systems-on-chip (SoCs)\rev{, such as Samsung Exynos and Nvidia\textregistered~Xavier\texttrademark}, combine the flexibility benefits of general-purpose cores with the energy efficiency and performance of custom designs. An emerging example is domain-specific SoCs, which integrate hardware accelerators targeting the commonly encountered tasks (i.e., computational kernels) in the target domain~\cite{hennessy2019new}. 

DSSoCs present a new challenge to the classical scheduling problem due to their specialized pipelines that run domain-specific tasks in the order of nanoseconds, i.e., orders of magnitude faster than general-purpose cores~\cite{hennessy2019new}.
Hence, achieving high performance with DSSoCs requires task scheduling algorithms that can execute in the order of nanoseconds.
Fast and low-overhead scheduling is an effective way to minimizing performance and energy consumption overheads. 
\rev{However, while enabling fast decision-making, simple schedulers can make poor scheduling decisions, especially under heavy workloads.} 

At low data rates, a low-overhead (fast) scheduler outperforms a more sophisticated scheduler due to the simplicity of the scheduling problem.
\rev{The number of concurrent tasks and the complexity of scheduling decisions grow with the data rate (heavy workload).} 
Consequently, the overhead of making better decisions pays off, i.e., the sophisticated scheduler starts outperforming the simple one. 
Hence, there is an opportunity to exploit the tradeoff between the scheduling overhead and decision quality.

We propose a \textit{dynamic adaptive scheduling} (DAS) 
framework that combines the benefits of both worlds, i.e., a simple scheduler with fast decision making and a sophisticated scheduler with high-quality scheduling decisions
through an integrated decision support mechanism. 
Making a scheduling decision at the scale of nanoseconds is highly challenging since it requires \rev{a scheduler to load the relevant feature data and execute} possibly complex decision criteria at the scale of nanoseconds. 
The following \textit{key observations} enable us to design the DAS framework that outperforms both types of schedulers taken separately:
\textit{First}, the scheduling is not an ordinary process that may be called in the future with some probability. Instead, it will be called with 100\% certainty and use a subset of available performance counters, i.e., features used for scheduling. Hence, a background process prefetches the relevant features and writes them to a pre-allocated local memory location. 
\textit{Second}, the same process can also determine whether a simple or a sophisticated scheduler with a higher overhead would perform better. 
If the lookup table (LUT) is preferred as the simple scheduler, 
the only extra delay on the critical path is the time it takes to access the LUT, which is 6 ns measured on Arm Cortex-A53.  
We run the sophisticated scheduler only if a complex decision is required at runtime.

\vspace{1mm}
The major contributions of this work are as follows:
\vspace{-1mm}
\begin{itemize}
    \item DAS framework that dynamically combines two 
    schedulers and outperforms each of them taken separately;
    \item \new{Low scheduling overhead: 4.2 nJ energy and 6 ns runtime for low to medium loads; 27.2 nJ energy and 65 ns runtime for heavy workloads};
    \vspace{-2mm}
    \item Experimental results with five streaming applications and profiling of scheduling overheads on a Xilinx Zynq ZCU102.
\end{itemize}

\section{Related Work}
\label{sec:related_work}

A variety of task scheduling algorithms, both static~\cite{topcuoglu2002performance}
and dynamic~\cite{pabla2009completely,krishnakumar2020runtime} have been proposed in the literature.
While the approaches in~\cite{topcuoglu2002performance, bittencourt2010dag} optimize the makespan of applications, completely fair scheduler (CFS) widely used in Linux-based OS aims to provide resource fairness to all processes~\cite{pabla2009completely}.
Although this model works well for the client and small-server systems, 
DSSoCs demand a novel suite of efficient schedulers that execute with nanosecond-scale overheads.
The scheduler complexity and overhead are discussed in~\cite{chronaki2016task,zhou2020real,namazi2018cmv}.
Chronaki \textit{et al.}~\cite{chronaki2016task} propose two dynamic schedulers, named CATS and CPATH, that detect the longest and critical paths in the application, respectively~\cite{chronaki2016task}. 
The benefits of the CPATH algorithm are rendered ineffective because of its higher scheduling overhead.

Scheduling algorithms proposed in~\cite{streit2002self} combine the benefits of using multiple schedulers.
\rev{They dynamically switch} between three schedulers to adapt to varying job characteristics.
However, the overheads of switching between policies are not considered when measuring scheduling overhead.
\rev{A performance discussion of a round-robin scheduler (simple, low complexity) and the earliest deadline first (high complexity) schedulers and their applicability under different system load scenarios are discussed in~\cite{zhou2020real}.}
In contrast, we combine the low scheduling overhead of a simple scheduler and the decision quality of a sophisticated scheduler based on the system workload.
To the best of our knowledge, this is the first approach that uses a novel runtime preselection classifier to choose between simple and sophisticated schedulers at runtime and thus enable nanosecond-scale overhead. 


\section{Dynamic Adaptive Scheduling Framework}
\label{sec:methodology}

\renewcommand{\arraystretch}{0.9}%
\begin{table}[b]
\centering
\vspace{1mm}
\caption{Type of performance counters used by DAS framework}
\label{tab:features}
\begin{tabular}{cc}
\toprule
\textbf{Type} & \textbf{Features}                                                                        \\ \midrule
Task         & \begin{tabular}[c]{@{}c@{}}Task ID, Execution time, Power consumption,\\ Depth of task in DFG, Application ID, \\Predecessor task ID and cluster IDs, Application type
\end{tabular}  \\ \midrule
\begin{tabular}[c]{@{}c@{}}Processing\\Element \\(PE)\end{tabular}           & \begin{tabular}[c]{@{}c@{}} Earliest time when PE is ready to execute,\\ Earliest availability time of each cluster, \\ PE utilization, Communication cost\end{tabular}                         \\ \midrule
System       & Input data rate     \\
\bottomrule
\end{tabular}
\end{table}

\subsection{Overview and Preliminaries} \label{sec:overview}
This work considers streaming applications that can be modeled by a 
data flow graph (DFG). 
Consecutive 
data frames are pipelined through the tasks in the graph. 
Unlike the current practice, which is limited to a single scheduler, DAS allows the OS to choose one scheduling policy $\pi \in \boldsymbol{\Pi_S} = \{F,S\}$, 
where $F$ and $S$ refer to the \textit{fast} and \textit{slow (or sophisticated)} schedulers, respectively.
Once the predecessors of a task are completed, the OS can call either a fast ($\pi=F$) or a slow scheduler ($\pi=S$) as a function of the system state and workload.
The OS collects a set of performance counters during the workload execution to enable two aspects for the DAS framework: (1) precise assessment of the system state, (2) desirable features for the classifier to dynamically switch between the \textit{fast} and \textit{slow} schedulers.

Table~\ref{tab:features} presents the 
performance counters collected by DAS.
For a DSSoC with 19 PEs, 
it uses 62 performance counters.
The goal of the fast scheduler $F$ is to approach the theoretically minimum (i.e., zero) scheduling overhead by making decisions in a few cycles with a minimum number of operations.
In contrast, the slow scheduler $S$ aims to handle more complex scenarios when the task wait times dominate the execution times. 
\textit{The goal of DAS is to outperform the optimization metrics (execution time and EDP) of both underlying schedulers by dynamically switching between them as a function of system state and workload.}

\subsection{Zero-Delay DAS Preselection Classifier} \label{sec:classifier}
\rev{The first step of DAS is selecting the fast or slow scheduler.} 
Since this decision is on the critical path of the fast scheduler, 
we must optimize it to approach our zero overhead goal. 
One of the novel contributions of DAS is recognizing this selection as a deterministic task that will eventually be  executed with probability one. 
Hence, we prefetch the relevant features required for this decision to a pre-allocated local register.
\rev{To minimize the overhead, we re-use a subset of the performance counters 
shown in Table~\ref{tab:features}
to make this decision, discussed in Section~\ref{subsec:model_methods}.
} 

The OS periodically refreshes the performance counters \rev{to reflect the current system state}.
Each time the features are refreshed, DAS runs a lightweight classifier that determines if the fast or slow scheduler should be used for the next ready task. 
\rev{This decision will always be up to date since it is refreshed with the features that reflect the most recent system state.}
This way, DAS determines which scheduler should be called even before a task is ready for scheduling. 
Hence, the preselection classifier introduces zero latency and minimal energy overhead, as described next.


\begin{figure}[!b]
	\centering
 	\includegraphics[draft=false,width=0.88\linewidth]{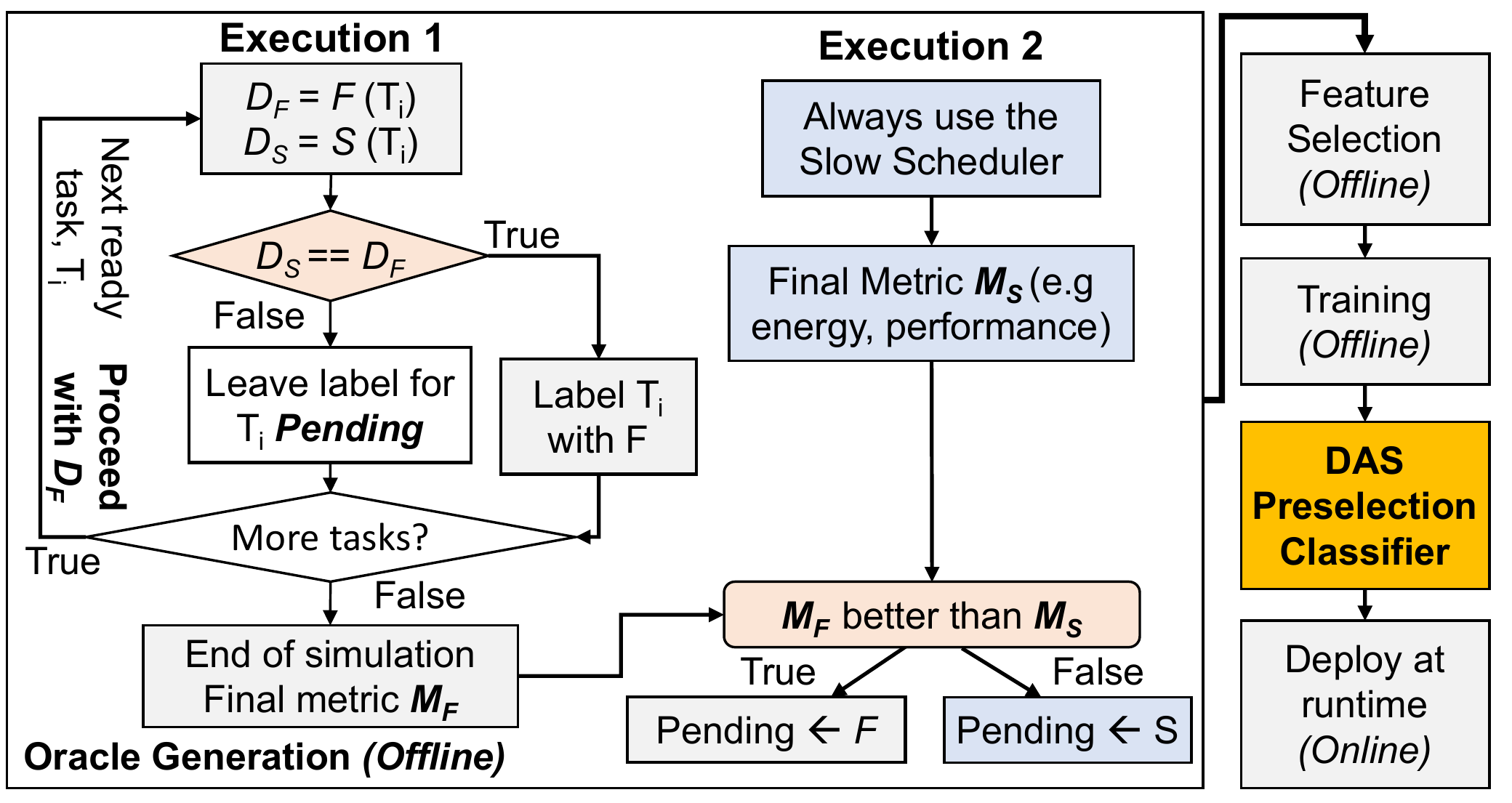}
	\caption{\rev{Flowchart describing the flow of the DAS framework: Oracle generation, feature selection, and training a model for the classifier.}}
	\label{fig:DAS_Oracle}
\end{figure}

\noindent \textbf{Offline Classifier Design:}
The first step to design the preselection classifier is generating the training data based on the domain applications known at design time. 
Each scenario in the training data consists of concurrent applications and their respective data rates 
(e.g., a combination of WiFi transmitter and receiver chains, at a specific upload and download speed). 
\rev{To this end, we run each scenario twice, as described in Figure~\ref{fig:DAS_Oracle}.}

\noindent\textit{First Execution:}
The instrumentation enables us to run \textit{both fast and slow schedulers} each time a task scheduling decision is made.
If the decisions of the fast ($D_F$) and slow ($D_S$) schedulers for a task $T_i$ are identical, then 
we label 
task $T_i$ with $F$ (i.e., the \textit{fast scheduler}) and store a
snapshot of the performance counters. 
If the schedulers return different decisions, then the label is left as \textit{pending}, and the execution continues by following the fast scheduler's decision, \rev{as illustrated in Figure~\ref{fig:DAS_Oracle}.
At the end of the first execution, the training data contains a mixture of both labeled ($F$) and pending decisions.}

\noindent\textit{Second Execution:}
The same scenario is executed, \textit{this time by always following the slow scheduler's decisions}. 
At the end of the execution, we analyze the target metric, \rev{such as the average execution time and energy-delay product.} 
If the slow scheduler achieves a better result, 
the pending labels are replaced with $S$ to indicate that the slow scheduler is preferred despite its larger overhead.
Otherwise, we conclude that the fast scheduler's lower overhead pays off and replace the pending labels with $F$. 
\rev{An alternative to replacing all pending labels at once is evaluating each decision individually. However, this approach will not be scalable since the scheduling decision at time $t_k$ affects not only the immediate action but also all the remaining execution flow.}

\rev{The training data is generated using 40 different workloads.} 
Each workload is a mix of multiple instances of five applications, consisting of approximately 140,000 tasks in total and executed at 14 different data rates (Section~\ref{subsec:sch_time_profiling}).
A higher data rate presents a larger number of concurrent applications contending for the same SoC resources.
Then, we design a low-overhead classifier using machine learning techniques and feature selection methods, as described in Section~\ref{subsec:model_methods} \rev{and shown in Figure~\ref{fig:DAS_Oracle}.} 


\noindent \textbf{Online Use of the Classifier:}
At runtime, a background process periodically updates a pre-allocated local memory with 
\rev{\textit{a small subset of performance counters} 
required by the classifier.}
After each update, the classifier determines whether the fast \textit{F} or slow \textit{S} scheduler should be used for the next available task.
When a new ready task becomes available, the features are already loaded, and we know which scheduler is a better choice. 
Therefore, DAS does not incur any extra delay on the critical path. Moreover, it has a negligible energy overhead, as demonstrated in Section~\ref{sec:experimental_results}.

\subsection{Fast \& Slow (Sophisticated) (F\&S) Schedulers} \label{sec:FSschedulers}

The DAS framework can work with any choice of fast and slow scheduling algorithms. 
This work uses a LUT implementation as the fast scheduler 
since the goal of the fast scheduler is to achieve almost zero overhead. 
The LUT stores the most energy-efficient processor in the target system for each known task in the target domain. Unknown tasks are mapped to the next available CPU core. 
Hence, the only extra delay on the critical path and overhead is the LUT access. 
To profile the scheduling overhead, we developed an optimized C implementation with inline assembly code. 
\rev{Experiments show that \textit{our fast scheduler takes $\sim$7.2 cycles (6 ns on Arm Cortex-A53 at 1.2~GHz) on average and incurs negligible (2.3 nJ) energy overhead.}}

The DAS framework uses a commonly used heuristic, earliest task first (ETF), as the slow scheduler~\cite{arda2020ds3}. 
\rev{ETF is chosen since it performs a comprehensive search to make a decision when the SoC is loaded with many tasks. 
It recursively iterates over the ready tasks and processors to find the schedule with the fastest finish time, as shown in Algorithm~\ref{algo:etf}. 
Hence, its computational complexity is quadratic on the number of ready tasks.} 



\begin{algorithm}[tb]
\footnotesize
\SetNoFillComment
\caption{ETF Scheduler}
\label{algo:etf}
\SetAlgoLined
\While {ready queue $\mathcal{T}$ is not empty} {
\For {task $T_i$ $\in \mathcal{T}$} {
\For {PE $p_j$ $\in \mathcal{P}$ \tcc{\textbf{$\mathcal{P}$ = set of PEs}} }  {
$FT_{T_{i},p_{j}}$ = Compute the finish time of $T_i$ on $p_j$  \\
}
}
 
($T'$, $p'$) = Find the task \& PE pair that has the minimum FT \\
Assign task $T'$ to PE $p'$
}
\end{algorithm}

\section{Experimental Results}
\label{sec:experimental_results}

\subsection{Experimental Setup}
\label{subsec:sch_time_profiling}
\noindent\textbf{Domain Applications:} The DAS framework is evaluated using five real-world streaming applications: 
range detection, temporal mitigation, WiFi-transmitter, WiFi-receiver applications, and a proprietary industrial application (\textit{App-1})~\cite{arda2020ds3,mack2020user}.
We construct 40 different workloads \rev{by mixing applications in different ratios for our evaluations. More information is provided in our github release for reproducibility~\cite{arda2020ds3}.}

\noindent\textbf{Emulation Environment:} One of our key goals in this study is to conduct a realistic energy and runtime overhead analysis. For this purpose, we leverage 
an open-source Linux-based emulation framework~\cite{mack2020user}.
For our analysis, we incorporate LUT and ETF schedulers into this emulation environment. 
We generate a wide range of workloads -- ranging from all application instances belonging to a single application to a uniform distribution from all five applications.
We measure the trend between the number of tasks ready to be scheduled and the scheduling overhead of ETF on the Xilinx Zynq ZCU102.
Based on these measurements, we generate a quadratic equation to formulate the ETF scheduling overhead. 
Later, we utilize this equation to evaluate the average execution time and the EDP of the DAS scheduler. 


\begin{figure*}[t!]
	\centering
 	\includegraphics[draft=false,width=0.78\linewidth]{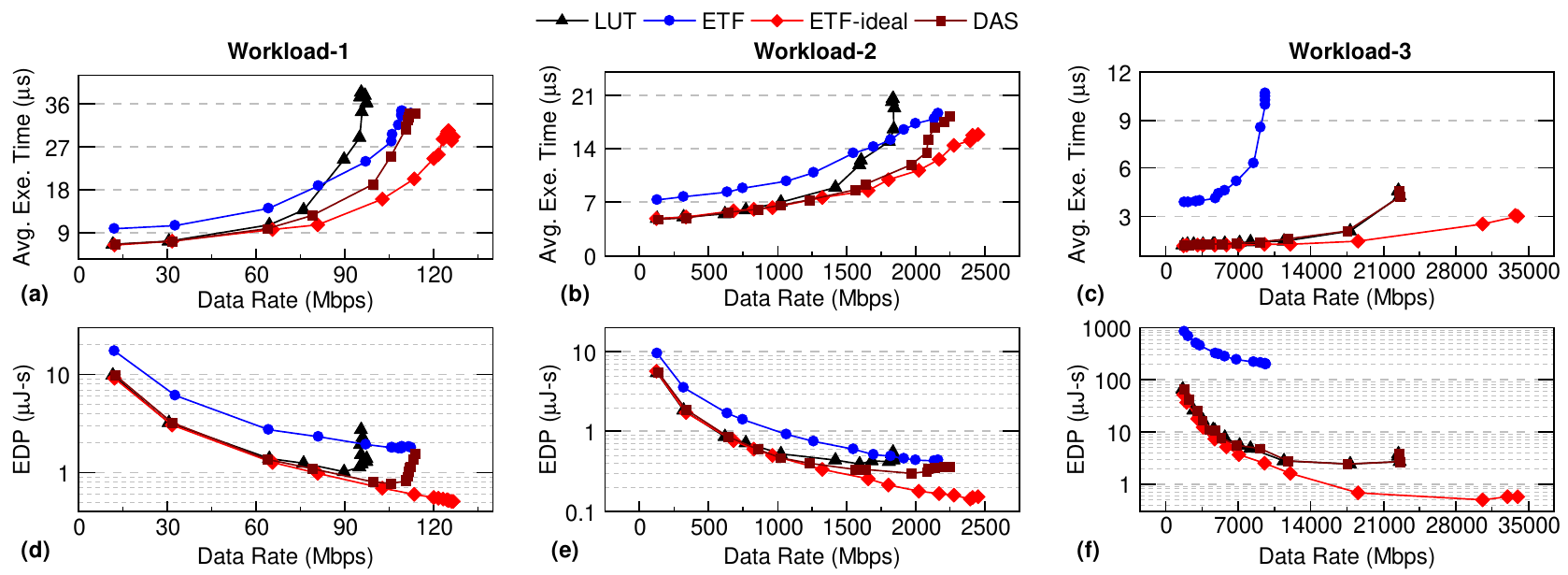}
 	\vspace{-1mm}
	\caption{Comparison of average execution time (a--c) and EDP (d--f) between DAS, LUT, ETF, and ETF-ideal for three different workloads.}
	\vspace{-1mm}
	\label{fig:exe_and_edp}
\end{figure*}

\noindent\textbf{Simulation Environment:} We use DS3~\cite{arda2020ds3}, an open-source domain-specific system-on-chip simulation framework, for the detailed evaluation of DAS.
DS3 includes built-in scheduling algorithms, models for PEs, interconnect, and memory systems.
The framework has been validated with Xilinx Zynq ZCU102 and Odroid-XU3 platforms.

\noindent\textbf{DSSoC Configuration:} We construct a DSSoC configuration that comprises clusters of general-purpose cores and hardware accelerators.
The application domains used in this study are wireless communications and radar systems.
The DSSoC used in our experiments uses the Arm big.LITTLE architecture with 4 cores each.
We also include dedicated accelerators for fast Fourier transform (FFT), forward error correction (FEC), finite impulse response (FIR), and a systolic array processor (SAP).
We include 4 cores each for the FFT and FIR accelerators, one core for the FEC, and two cores of the SAP.
The FEC accelerates the execution of encoder and decoder operations.
In total, the DSSoC integrates 19 PEs with a mesh-based network-on-chip to enable efficient on-chip data movement.


\subsection{Exploration of Machine Learning Techniques and Feature Space  for DAS}
\label{subsec:model_methods}

\noindent \textbf{Machine Learning Technique Exploration:} We explore different classifiers to co-optimize the \textit{classification accuracy}
and 
\textit{model size} 
towards our minimal overhead goal.
Specifically, we investigated support vector classifiers, decision tree (DT), multi-layer perceptron (MLP), and logistic regression (LR).
The training process with support vector classifiers with simple kernels exceeded 24 hours, rendering it infeasible.
The latency and storage requirements of the MLP (one hidden layer and 16 neurons) did not fit the budgets of low-overhead requirements.
Therefore, these two techniques are excluded from the rest of the analysis.
Table~\ref{tab:classifiers} summarizes the classification accuracy and storage overheads for the LR and DT classifiers as a function of the number of features.
%
DTs achieve similar or higher accuracies compared to 
LR classifiers with lower storage overheads. 
While a DT with depth 16 
that uses all features achieves the best classification accuracy, there is a significant impact on the storage overhead, which in turn influences the latency and energy consumption of the classifier.
In comparison, DTs with depth 2 and 4 have negligible storage overheads with competitive accuracies (>85\%). 
Hence, for the DAS framework, we adopt the DT classifier with depth 2.

\begin{table}[b]
\centering
\caption{Classification accuracies and storage overhead of DAS models with different machine learning classifiers and features}
\label{tab:classifiers}
\begin{tabular}{@{}ccc|cc@{}}
\toprule
\textbf{Classifier} & \textbf{Tree Depth} & \textbf{\begin{tabular}[c]{@{}c@{}}Number of \\ Features\end{tabular}} & \textbf{\begin{tabular}[c]{@{}c@{}}Classification\\ Accuracy (\%)\end{tabular}}  & \textbf{\begin{tabular}[c]{@{}c@{}}Storage\\ (KB)\end{tabular}} \\ \midrule
\textbf{LR}         & -  & 2  & 79.23                        & 0.01         \\
\textbf{LR}         & -  & 62 & 83.1                         & 0.24         \\
\textbf{DT}         & 2  & \textbf{1}  & \textbf{63.66}                        & 0.01        \\ 
\textbf{DT}         & 2  & \textbf{2}  & \textbf{85.48}                        & 0.01        \\ 
\textbf{DT}         & 4  & 6  & 85.51                        & 0.03        \\ 
\textbf{DT}         & 16 & 62 & 91.65                        & 256         \\
\bottomrule
\end{tabular}
\end{table}

\noindent \textbf{Feature Space Exploration:}
We collect 62 performance counters in our training data. 
A systematic feature space exploration is performed using feature selection and importance methods.
\rev{Among the top six features, growing the feature list from a single feature (\emph{input data rate}) to two features with the addition of \emph{the earliest availability time of the Arm big cluster} increases the accuracy from 63.66\% to 85.48\%.}
The data rate is tracked at runtime by an 8-entry$\times$16-bit shift register.
Therefore, we utilize \rev{only} two most important features to design a DT of depth 2 for the DAS classifier model\rev{; this takes 13 ns to execute on Arm Cortex-A53 cores running at 1.2 GHz.}
\subsection{Performance and Scheduling Overhead Analysis} \label{subsec:workloads}


This section compares the DAS framework with LUT (\textit{fast}), ETF (\textit{slow}),
and ETF-ideal schedulers.
ETF-ideal is a version of the ETF scheduler which ignores the scheduling overhead. It helps us establish the theoretical limit of achievable execution time and EDP.
Out of the 40 workloads described in Section~\ref{sec:classifier}, we choose \rev{three representative workloads for a detailed analysis of execution time and EDP trends.
These workloads} present different data rates, which are a function of the applications in the workload.
Workload-1 (Figures~\ref{fig:exe_and_edp}a,d) presents low data rate, workload-2 (Figures~\ref{fig:exe_and_edp}b,e) presents moderate data rates, and workload-3 (Figures~\ref{fig:exe_and_edp}c,f) represents a high data rate workload.

\begin{figure}[b]
	\centering
 	\includegraphics[draft=false,width=0.89\linewidth]{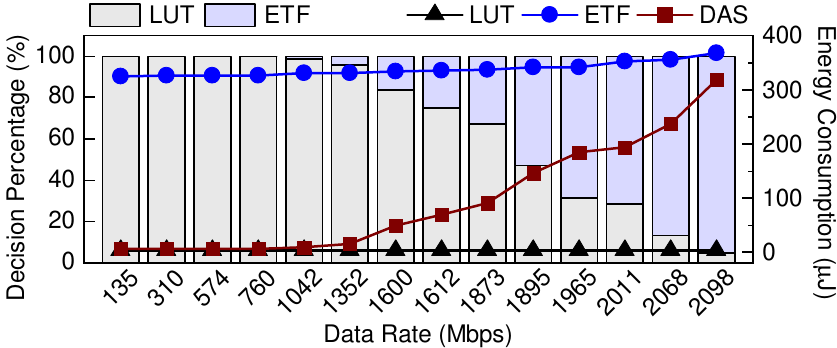}
	\caption{Decisions taken by the DAS framework as bar plots and total scheduling energy overheads of LUT, ETF, and DAS as line plots.}
	\label{fig:decision_dist}
\end{figure}

Figures~\ref{fig:exe_and_edp}a--c (Figures~\ref{fig:exe_and_edp}d--f) compare the execution times (EDP) of DAS, LUT, ETF, and ETF-ideal.
\rev{For workloads 1 and 2, the SoC is not congested at low data rates. Hence, DAS performs similar to LUT.
As data rates increase, DAS aptly chooses between LUT and ETF at runtime. Its execution time and EDP is 14\% and 15\% lower than LUT, and 15\% and 42\% lower than ETF.}
For workload-3, the execution time and EDP of ETF are significantly higher than LUT.
DAS chooses LUT for >99\% of the decisions and closely follows its execution time and EDP.

\rev{This} study is extended to all 40 workloads. 
At low data rates, DAS achieves 1.29$\times$ speedup and 45\% lower EDP compared to ETF, and 1.28$\times$ speedup and 37\% lower EDP than LUT, when the workload complexity increases. 
In summary, DAS \rev{consistently performs better than either one of the underlying schedulers, successfully adapts to the workloads at runtime, and aptly chooses between LUT and ETF to achieve low execution time and EDP.}

The \rev{left} axis of Figure~\ref{fig:decision_dist} plots the decision distribution of DAS. 
It uses LUT for \textit{all} decisions at the lowest data rate and ETF for 95\% of decisions at the highest data rate.
At a moderate workload of 1352 Mbps, DAS still uses LUT for 96\% of the decisions. 
The secondary axis of Figure~\ref{fig:decision_dist} shows the energy overhead of using different schedulers. 
As DAS uses LUT and ETF based on the system load, its energy consumption varies from that of LUT to ETF. 
The average scheduling latency overhead of DAS under heavy workloads is 65 ns, and the energy overhead is 27.2 nJ. 

\rev{We also compared DAS against a heuristic that chooses the fast scheduler when the data rate is less than a predetermined threshold and uses the slow scheduler otherwise. The threshold is chosen judiciously by analyzing the training data used for DAS. Simulation results show that the heuristic closely follows LUT (fast) and ETF (slow) schedulers below and above the data rate threshold, respectively. In contrast, DAS consistently outperforms both schedulers and achieves on average 13\% lower execution time than the heuristic across all data rates.}




\section{Conclusion}
\label{sec:conclusion}

In this paper, we presented a dynamic adaptive scheduling framework that combines the benefits of \emph{fast} and \emph{sophisticated} schedulers 
for heterogeneous SoCs.
DAS achieves an overhead that is as low as 6 ns (4.2 nJ) for a wide range of workload scenarios and on average, 65 ns (27.2 nJ) for heavy workloads for wireless communication and radar system applications.
Hence, our approach paves the way for DSSoCs to leverage their potential better to enable peak performance and energy-efficiency of domain applications.



\bibliographystyle{IEEEtran}
\footnotesize{\bibliography{main}}

\begin{thebibliography}{10}
\providecommand{\url}[1]{#1}
\csname url@samestyle\endcsname
\providecommand{\newblock}{\relax}
\providecommand{\bibinfo}[2]{#2}
\providecommand{\BIBentrySTDinterwordspacing}{\spaceskip=0pt\relax}
\providecommand{\BIBentryALTinterwordstretchfactor}{4}
\providecommand{\BIBentryALTinterwordspacing}{\spaceskip=\fontdimen2\font plus
\BIBentryALTinterwordstretchfactor\fontdimen3\font minus
  \fontdimen4\font\relax}
\providecommand{\BIBforeignlanguage}[2]{{%
\expandafter\ifx\csname l@#1\endcsname\relax
\typeout{** WARNING: IEEEtran.bst: No hyphenation pattern has been}%
\typeout{** loaded for the language `#1'. Using the pattern for}%
\typeout{** the default language instead.}%
\else
\language=\csname l@#1\endcsname
\fi
#2}}
\providecommand{\BIBdecl}{\relax}
\BIBdecl

\bibitem{hennessy2019new}
J.~L. Hennessy \emph{et~al.}, ``{A New Golden Age for Computer Architecture},''
  \emph{Commun. of the ACM}, vol.~62, no.~2, pp. 48--60, 2019.

\bibitem{topcuoglu2002performance}
H.~Topcuoglu \emph{et~al.}, ``{Performance-Effective and Low-Complexity Task
  Scheduling for Heterogeneous Computing},'' \emph{IEEE Trans. on Parallel and
  Distrib. Syst.}, vol.~13, no.~3, pp. 260--274, 2002.

\bibitem{pabla2009completely}
C.~S. Pabla, ``{Completely Fair Scheduler},'' \emph{Linux Jrnl.}, no. 184,
  2009.

\bibitem{krishnakumar2020runtime}
A.~Krishnakumar \emph{et~al.}, ``{Runtime Task Scheduling using Imitation
  Learning for Heterogeneous Many-core Systems},'' \emph{IEEE Trans. on CAD of
  Integr. Circuits and Syst.}, vol.~39, no.~11, pp. 4064--4077, 2020.

\bibitem{bittencourt2010dag}
L.~F. Bittencourt \emph{et~al.}, ``{DAG Scheduling Using a Lookahead Variant of
  the Heterogeneous Earliest Finish Time Algorithm},'' in \emph{IEEE Euromicro
  Conf. on Parallel, Distrib. and Network-based Process.}, 2010, pp. 27--34.

\bibitem{chronaki2016task}
K.~Chronaki \emph{et~al.}, ``{Task Scheduling Techniques for Asymmetric
  Multi-core Systems},'' \emph{IEEE Trans. on Parallel and Distrib. Systems},
  vol.~28, no.~7, pp. 2074--2087, 2016.

\bibitem{zhou2020real}
J.~Zhou, ``{Real-time Task Scheduling and Network Device Security for Complex
  Embedded Systems based on Deep Learning Networks},'' \emph{Microprocessors
  and Microsystems}, vol.~79, 2020.

\bibitem{namazi2018cmv}
A.~Namazi \emph{et~al.}, ``{CMV: Clustered Majority Voting Reliability-aware
  Task Scheduling for Multicore Real-time Systems},'' \emph{IEEE Trans. on
  Reliability}, vol.~68, no.~1, pp. 187--200, 2018.

\bibitem{streit2002self}
A.~Streit, ``{A Self-tuning Job Scheduler Family with Dynamic Policy
  Switching},'' in \emph{Workshop on Job Scheduling Strategies for Parallel
  Process.}\hskip 1em plus 0.5em minus 0.4em\relax Springer, 2002, pp. 1--23.

\bibitem{arda2020ds3}
S.~E. Arda \emph{et~al.}, ``{DS3: A System-Level Domain-Specific System-on-Chip
  Simulation Framework},'' \emph{IEEE Trans. on Computers}, vol.~69, no.~8, pp.
  1248--1262, 2020, [Online] \url{https://github.com/segemena/DS3.git}.

\bibitem{mack2020user}
J.~Mack \emph{et~al.}, ``{User-Space Emulation Framework for Domain-Specific
  SoC Design},'' in \emph{Proc. IPDPS Workshops}, 2020, pp. 44--53.

\end{thebibliography}

\end{document}